# Influence de la teneur en protéines de solutions physiologiques sur le comportement électrochimique du Ti-6Al-4V : reproductibilité et représentation temps-fréquence.

Jean Geringer[a], Laurent Navarro[a], Bernard Forest[a]

[a] *Ecole Nationale Supérieure des Mines de Saint-Etienne, ENSM-SE*
*Centre Ingénierie Santé, Département B2M*
*UMR CNRS PECM 5146, IFR 143*
*158, cours Fauriel F-42023 Saint-Etienne*
*Tel: (33).4.77.42.66.88,*
*E-mail : geringer@emse.fr ; navarro@emse.fr ; forest@emse.fr*

**Résumé :** Le comportement électrochimique des alliages métalliques biomédicaux, notamment dans le domaine des implants orthopédiques, pose encore de nombreuses questions. Ce travail propose d'étudier l'alliage de titane Ti-6Al-4V, par spectroscopie d'impédance électrochimique, SIE, dans différents milieux physiologiques : solution de Ringer, solution à base d'un tampon phosphate (PBS), solution PBS avec de l'albumine, solution PBS avec du sérum bovin et une solution PBS avec du sérum bovin et un antioxydant (azoture de sodium). De plus, une solution d'eau ultra-pure servira de référence. La reproductibilité des tests a été étudiée. Les représentations temps-fréquence des modules ont mis en évidence que l'eau désionisée est la solution qui présente le caractère le plus protecteur pour le Ti-6Al-4V. Cet alliage de titane est le moins protégé dans la solution de PBS contenant de l'albumine. Cette représentation permet de mettre en évidence des signatures graphiques d'adsorption des espèces inorganiques et organiques (différences entre les moyennes des modules dans les solutions étudiées et la moyenne des modules dans la solution de référence).

**Mots clés :** alliage de titane; biomatériau; protéines; représentation temps-fréquence

# Influence of proteins from physiological solutions on the electrochemical behaviour of the Ti-6Al-4V alloy: reproducibility and time-frequency dependence.

**Abstract:** The electrochemical behaviour of the biomedical and metallic alloys, especially in the orthopaedic implants fields, raises many questions. This study is dedicated for studying the Ti-6Al-4V alloy, by electrochemical impedance spectroscopy, EIS, in various physiological media,: Ringer solution, phosphate buffered solution (PBS), PBS solution and albumin, PBS solution with calf serum and PBS solution with calf serum and an antioxidant (sodium azide). Moreover, the desionised water was considered as the reference solution. The tests reproducibility was investigated. The time-frequency-Module graphs highlighted that the desionised water is the most protective for the Ti-6Al-4V alloy. This biomedical alloy is the less protected in the solution constituted by PBS and albumin. The time-frequency graph allows pointing out the graphic signatures of adsorption for organic and inorganic species (differences between the modules means in studied solution and the modules mean in the reference solution).

# 1. Introduction

Les matériaux métalliques, utilisés comme implants orthopédiques et notamment dans les prothèses de hanche, sont : l'alliage de titane, Ti-6Al-4V, l'acier inoxydable 316L voire 316LN et l'alliage de Co-Cr-Mo à hautes et basses teneurs en carbone [1]. Après leur implantation et en raison des interactions avec le liquide physiologique et les tissus, de nombreuses questions restent sans réponse au sujet des interactions physico-chimiques entre la surface du biomatériau, les ions phosphates (tampon phosphate présent dans le liquide physiologique) et les protéines.

Des investigations sont en cours pour comprendre les interactions des substances inorganiques et organiques avec les biomatériaux métalliques. La passivité de l'alliage de Co-Cr-Mo en présence de protéines, sérum bovin (milieu riche en protéines dont l'albumine) ou albumine, et/ou les ions phosphates est le sujet de nombreuses expériences par spectroscopie d'impédance électrochimique (SIE). Tous les résultats issus de la bibliographie qui seront présentés par la suite ont été obtenus grâce à l'utilisation d'une solution thermostatée à 37 °C. Hodgson et al [2], dans une solution contenant des ions phosphates et non des protéines, présente que le film d'oxydes de l'alliage Co-Cr-Mo s'enrichit en Cr lorsque de tels échantillons sont en contact avec des solutions dont la composition en espèces inorganiques est proche du liquide physiologique. De plus, le cobalt se dissout préférentiellement [2].

Dans une solution contenant des ions phosphates et du sérum bovin, il a été mis en évidence que les ions phosphates bloquent cinétiquement les réactions de corrosion de l'alliage de Co-Cr-Mo ; la résistance de transfert de charge augmente lorsque la concentration en ions phosphates augmente. La présence de sérum bovin n'a pas d'influence significative sur la corrosion de l'alliage. Le modèle électrique équivalent, 2 circuits RC en parallèle, a été choisi [3].

La compétition de l'adsorption entre les ions phosphates et l'albumine, avec l'alliage de Co-Cr-Mo, a été étudiée par Munoz et al. [4]. Plus la concentration en ions phosphates augmente, plus la résistance à la corrosion de l'alliage augmente. D'après des études de spectroscopie d'impédance électrochimique, un circuit électrique équivalent composé de 2 circuits RC en série a été proposé pour décrire la surface de l'alliage. L'action de l'albumine n'a pas pu être mise en évidence clairement ; elle dépend fortement de l'état du film passif avant les tests.

Des investigations supplémentaires [5] ont complété les résultats précédents. Une comparaison a été menée entre l'alliage Co-Cr-Mo et l'acier inoxydable 316L. Pour les deux alliages, les ions phosphates agissent comme un inhibiteur anodique et l'albumine comme un inhibiteur cathodique. L'albumine diminue la résistance de transfert de charge pour le 316L et l'augmente pour le Co-Cr-Mo. Enfin, la durée d'immersion ne modifie pas le comportement du 316L. En revanche, plus la durée d'immersion augmente, plus la résistance à la corrosion de Co-Cr-Mo augmente avec la concentration en ions phosphates ; l'albumine entraîne une évolution opposée, la résistance à la corrosion diminue avec l'augmentation de la teneur en albumine. En conclusion, il semble que l'acier inoxydable 316L est moins sensible à la présence de protéines que le Co-Cr-Mo.

Ces résultats montrent que la caractérisation électrochimique de l'adsorption de l'albumine est difficile à mettre en évidence, en raison de l'importance de l'état de passivation de la surface de l'alliage.

Nous allons maintenant nous intéresser plus particulièrement à l'alliage de titane, Ti-6Al-4V, surtout utilisé pour fabriquer des tiges de prothèses de hanche ou des implants dentaires. Il est reconnu pour sa biocompatibilité, les tissus et les cellules se reconstruisent plus aisément sur sa surface que sur les surfaces des autres biomatériaux métalliques, et ses propriétés mécaniques qui lui permettent de répondre au cahier des charges de la fabrication d'une tige fémorale [6,7]. En revanche, le Ti-6Al-4V ne possède pas de bonnes propriétés tribologiques, notamment en milieu aqueux, problématique de tribocorrosion. Souvent, il est nécessaire de déposer un revêtement tel TiN pour améliorer sa résistance à la corrosion [8].

Une surface polie permet de diminuer sensiblement, 2 ordres de grandeur, le courant de corrosion [9]. En revanche, dans une solution de PBS, à 22 °C, le potentiel de corrosion est voisin de -300 mV/ECS et ne dépend pas e la rugosité de surface [9].

Ensuite, il a été montré que le liquide physiologique proche du liquide synovial, ponctionné d'articulations de patients, est beaucoup plus corrosif que du sérum ou de l'urine. De plus, l'augmentation de la température diminue la passivité du Ti-6Al-4V [10].

La double couche à la surface de cet alliage, dans une solution contenant des protéines à 37 °C, contient des ions hydroxylés. Il a été établi que le transfert de charges se produit à travers cette couche. Les protéines, quant à elles, favorisent la diffusion des cations métalliques dissous, d'où une accélération de la vitesse de corrosion, à pH inférieur à 7. Dans un domaine de pH compris entre 7 et 9, l'adsorption des protéines entraîne la formation d'une couche protectrice. Enfin, la corrosion de cet alliage entraîne la dégradation de la couche d'oxydes, rendant alors la surface plus poreuse ; la dureté diminue, confirmant cette hypothèse [11].

La surface du Ti-6Al-4V, plongé dans une solution simulant le liquide physiologique, possède une couche d'oxydes et une couche d'adsorbat, des protéines et/ou de l'apatite. A partir d'expériences de SIE, il a été clairement mis en évidence deux constantes de temps, [12,13]. La description de l'interface, par les modèles électriques équivalents, a conduit à proposer l'association de deux circuits RC en parallèle pour décrire l'interface entre le métal et la solution contenant des protéines et des ions phosphates [14]. Elle serait composée d'une couche d'oxydes, d'une couche d'espèces inorganiques (ions phosphates) et d'une couche de protéines issues de sérum bovin, à la surface du titane pur [15]. Une compétition existe entre l'adsorption des phosphates et des protéines, d'où les comportements, déterminés par spectroscopie d'impédance électrochimique qui peuvent être parfois antagonistes [14,15].

Enfin, pour terminer cette description, il est intéressant de noter que la quantité de protéines adsorbées à la surface du titane pur est comprise entre 1,8 et 2,89 µg.cm$^{-2}$, solution contenant des phosphates et des protéines à 37 °C [16,17]. La méthode de dosage des protéines adsorbées est indirecte. Elle est basée sur le dosage des protéines libres en solution par spectrophotométrie UV, après avoir mis une quantité connue à la surface de l'échantillon métallique. D'autres travaux ont montré que la quantité de BSA adsorbée à la surface d'un échantillon de Ti, à 37 °C, était égale à 710 ± 70 ng.cm$^{-2}$ [18]. La technique de mesure est basée sur l'utilisation d'une microbalance à quartz. Les teneurs varient d'un facteur 4 entre ces deux références. On peut penser que la première méthode, dosage indirect, est moins précise que la deuxième. Ensuite, l'état de surface des échantillons, rugosité moyenne par exemple, n'est pas précisé et pourrait jouer un rôle sur l'adsorption des protéines. De plus, la première série d'expériences concerne des dépôts de titane projetés sur des wafers de silicium. La deuxième série concerne des dépôts de titane sur des cristaux de quartz. Les conditions de dépôts n'assurent surement pas le même état de surface d'où une surface spécifique différente. La concentration, à la surface des échantillons de titane poli miroir (Ra < 0,05 µm), pourrait même être inférieure à celles énoncées plus haut ; la surface spécifique sera moins grande que dans le cas de dépôts de titane sur un substrat.

De façon à mieux comprendre le comportement de cet alliage de titane Ti-6Al-4V, nous proposons d'étudier l'évolution de la surface, par spectroscopie d'impédance électrochimique, dans différentes solutions d'étude de façon à appréhender l'influence des ions phosphates, de l'albumine et des protéines. Une représentation temps-fréquence sera proposée.

## 2. Matériaux et méthodes

### 2.1 Matériau testé

L'alliage de Ti-6Al-4V a la composition suivante, Tableau 1.

| Eléments | Al | V | Fe | O | Cr | Ni | C |
|---|---|---|---|---|---|---|---|
| Teneur | 5,91 % | 3,87 % | 1096 ppm | 1038 ppm | 137 ppm | 125 ppm | 124 ppm |

| Eléments | N | H | Co | Mo | Mn | Cu | B |
|---|---|---|---|---|---|---|---|
| Teneur | 53 ppm | 29 ppm | < 300 ppm | < 100 ppm | < 50 ppm | < 20 ppm | < 10 ppm |

*Tableau 1 : composition de l'alliage Ti-6Al-4V.*

*Table 1: chemical composition of the Ti-6Al-4V alloy.*

Une des faces des pastilles, épaisseur de 5 mm et diamètre de 35 mm, a été polie 'miroir' grâce à une gamme de polissage comprenant des papiers abrasifs de 240 jusqu'à 2400 et des solutions diamantées contenant des particules de 3 à 1 µm. Les échantillons ont été rincés avec de l'eau désionisée et séchés à l'air.

### 2.2 Solutions d'étude

Six solutions, 1- eau désionisée/2- solution de Ringer/3- solution de PBS (Phosphate Buffered Solution)/4- solution de PBS + Alb (Albumine)/5- solution PBS + sérum bovin/6- solution PBS + sérum bovin + $NaN_3$ (antioxydant), ont été considérées de façon à appréhender l'influence des ions phosphates, de l'albumine (Albumine fraction V extraite d'un sérum bovin, protéine 'modèle') et des protéines d'un sérum bovin (sérum bovin, calf serum heat inactivated, PAA, triple filtrage de porosité 0,1 µm), Tableau 2, sur le comportement passif de la surface de l'alliage Ti-6Al-4V. Le sérum bovin contient 74 $g.L^{-1}$ de protéines, concentration totale des protéines. Plus précisément, le sérum utilisé est constitué de 35 $g.L^{-1}$ d'albumine, 9 $g.L^{-1}$ de α-globuline, 11 $g.L^{-1}$ de β-globuline et 19 $g.L^{-1}$ de γ-globuline. Dans les solutions d'étude, le sérum bovin a été dilué par 4. Les compositions de ces solutions ont été proposées à partir de l'étude effectuée sur un alliage de Co-Cr-Mo, Round Robin dans le cadre du COST 533 [19].

| Concentration massique / g.L$^{-1}$ | NaCl | KCl | CaCl$_2$,2H$_2$O | NaHCO$_3$ | KH$_2$PO$_4$ | Na$_2$HPO$_4$ | Albumine | Sérum bovin | NaN$_3$ |
|---|---|---|---|---|---|---|---|---|---|
| Solution 1, eau désionisée | | | | | | | | | |
| Solution 2, solution Ringer | 8,500 | 0,250 | 0,220 | 0,150 | | | | | |
| Solution 3, solution PBS | 8,192 | 0,223 | | | 0,136 | 1,420 | | | |
| Solution 4, solution PBS + Alb | 8,192 | 0,223 | | | 0,136 | 1,420 | 1,000 | | |
| Solution 5, PBS + Sérum | 8,192 | 0,223 | | | 0,136 | 1,420 | | 18,5 | |
| Solution 6, PBS + Sérum + NaN$_3$ | 8,192 | 0,223 | | | 0,136 | 1,420 | | 18,5 | 1,000 |

*Tableau 2 : composition des six solutions utilisées, PBS : solution tampon phosphate (Phosphate Buffered Saline) ; Alb : Albumine ; sérum : sérum bovin ou calf sérum ; NaN$_3$ : azoture de sodium, antioxydant.*

*Table 2: composition of six investigated solutions, PBS: Phosphate Buffered Saline; Alb : Albumin ; sérum: calf serum ; NaN$_3$ : sodium azide, antioxidant.*

Les solutions ont été portées à une température de 37 ± 1 °C, durant les expériences, pour correspondre à la température du liquide physiologique réel.

*2.3 Les conditions expérimentales*

Le Tableau 3 présente les conditions expérimentales utilisées.

| | |
|---|---|
| *Marque du potentiostat* | *Parstat 2263* |
| *Taille électrode de travail, WE* | *7,07 cm2 ; diamètre de 30 mm* |
| *Electrode de référence, RE* | *ECS ; 0,250 V/ESH* |
| *Contre électrode, CE* | *Fil de platine, cercle de diamètre 25 mm* |
| *Distance RE-WE* | *10 ± 2 mm* |
| *Volume de la solution* | *60 ± 2 mL* |
| *pH de la solution* | *7,6 ± 0,1* |
| *Température* | *37 ± 1 °C* |
| *Dispositif thermostaté* | *Enceinte double enveloppe* |

*Tableau 3 : description des conditions expérimentales, ECS : Electrode au Calomel Saturé, ESH : Electrode Standard Hydrogène.*

*Table 3: experimental conditions, SCE (ECS): Saturated Calomel Electrode, SHE (ESH): Standard Hydrogen Electrode.*

*2.4 Les conditions électrochimiques*

Il a été montré dans l'étude bibliographique que l'état du film passif, pour étudier l'adsorption des ions phosphates et des protéines, était un facteur prépondérant. En conséquence, il a été décidé de polariser cathodiquement chaque échantillon, avant le début des mesures de spectroscopie d'impédance électrochimique. Quatre séquences ont été effectuées pour chaque essai :

1. Potentiel imposé – 1 V/ECS pendant 5 minutes
2. Potentiel libre de corrosion pendant 2 heures
3. Début d'une boucle, répétée 120 fois
    a. SIE, 100 kHz à 100 mHz, 20 points/décade, amplitude de ± 10 mV, Data quality de 15, mesures effectuées à potentiel libre
    b. Potentiel libre de corrosion durant 2 min
4. Potentiel libre de corrosion durant 26 heures.

Dans l'étape 3a., le paramètre Data quality modifie l'algorithme d'acquisition des données, en améliorant la qualité de cette acquisition. Si Data quality augmente, le nombre d'acquisitions augmente, pour un même point de mesure, ainsi la durée d'acquisition augmente mais le rapport signal sur bruit diminue.

Toutes les manipulations ont été répétées 3 fois, exactement dans les mêmes conditions. Le montage avec la double enveloppe a été inséré dans une cage de Faraday., i.e. toutes les connexions, les échantillons et toutes les électrodes.

## 3. Résultats et discussions

*3.1 Méthodes de comparaison des résultats*

La Figure 1 a) présente le diagramme de Nyquist obtenu pour un échantillon de Ti-6Al-4V immergé dans de l'eau désionisée. Le temps, 0 h, correspond au début de la 3$^{ème}$ séquence, i.e. après 2 heures à potentiel libre et est relatif à la première expérience de SIE ; le temps, 12,5 h, est relatif à la 120$^{ème}$ expérience de SIE. On constate que le comportement électrochimique de la surface de l'alliage de Ti-6Al-4V évolue au cours du temps. La boucle capacitive diminue au cours du temps. On peut considérer que la couche, à la surface du matériau, comptant pour l'interprétation du circuit électrique équivalent à hautes fréquences ($R_{solution}(C_{hf}R_{hf})$), est moins protectrice. La Figure 1 b) montre les modules et les phases correspondant à ces mêmes expériences. Dans le domaine fréquentiel 10-100 kHz, les valeurs de phase montrent que le comportement n'est pas purement résistif ; ces points correspondent à des points aberrants. Avec cette représentation, entre 10 et 100 kHz, les variations de module sont difficilement exploitables. On constate aussi que la phase diminue, en dessous de 1 Hz, pour la mesure à 12,5 h.

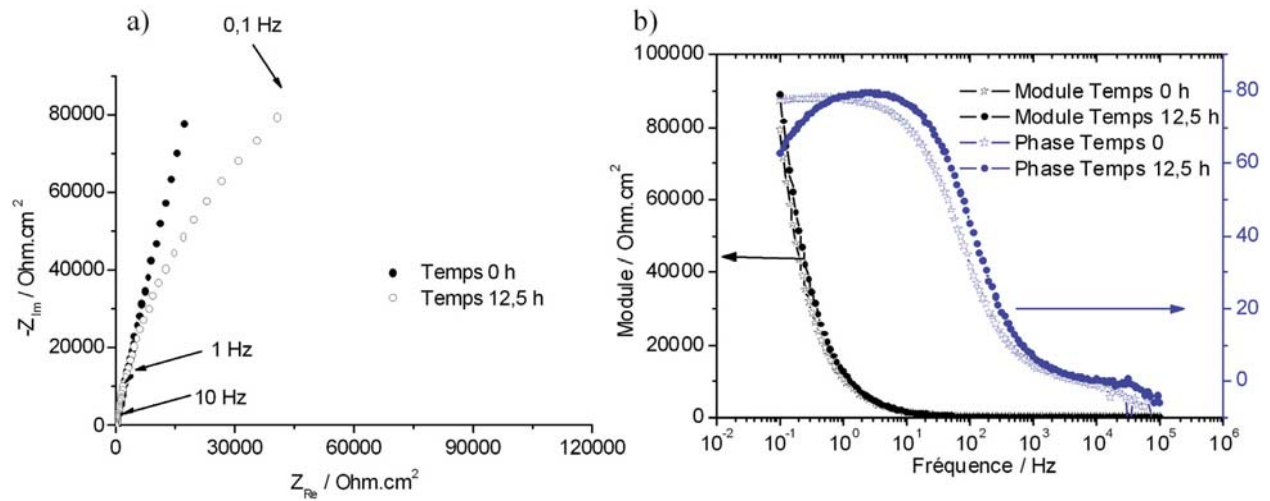

*Figure 1 : a) Diagramme de Nyquist correspondant à un échantillon de Ti-6Al-4V immergé dans de l'eau désionisée ; temps 0 h : 1$^{ère}$ expérience de SIE ; temps 12,5 h : 120$^{ème}$ expérience de SIE ; b) Modules et phases relatifs aux mêmes expériences.*

*Figure 1: a) Nyquist Diagram related to Ti-6Al-4V sample in desionised water; time 0 h: 1$^{st}$ SIE experience; time 12.5 h : SIE experience 120; b) Modules et phases related to 1$^{st}$ experience and experience 120.*

De façon à pouvoir conclure sur l'évolution de ce matériau, il est possible de calculer les valeurs du circuit électrique équivalent et de les comparer au cours du temps, par exemple tracer l'évolution de la résistance à hautes fréquences en fonction du temps. Une étude de reproductibilité est aussi nécessaire car l'état de surface, notamment la topologie de surface, ne peut pas être rigoureusement le même entre différents échantillons. L'intérêt de réaliser autant de mesures de SIE, i.e. 120, est d'avoir des informations sur la cinétique d'apparition des phénomènes compétitifs d'adsorption des espèces inorganiques, ions phosphates, et des espèces organiques, albumine et sérum bovin. De plus, la comparaison des résultats entre différentes solutions peut consister en l'examen des valeurs des constituants des circuits électriques équivalents, cette méthode est d'ailleurs très instructive. Néanmoins, une visualisation graphique n'est pas couramment utilisée. C'est la raison pour laquelle, une représentation 3D du module dans le plan temps-fréquence sera proposée, grâce à un programme Matlab® de traitement des données brutes. De la Figure 1 b), on peut représenter la Figure 2 a) qui, en projection dans le plan temps-fréquence, donne la Figure 2 b). Le module sera représenté en $\Omega$, par la suite ; la surface est égale à 7,07 cm$^{-2}$, tableau 3.

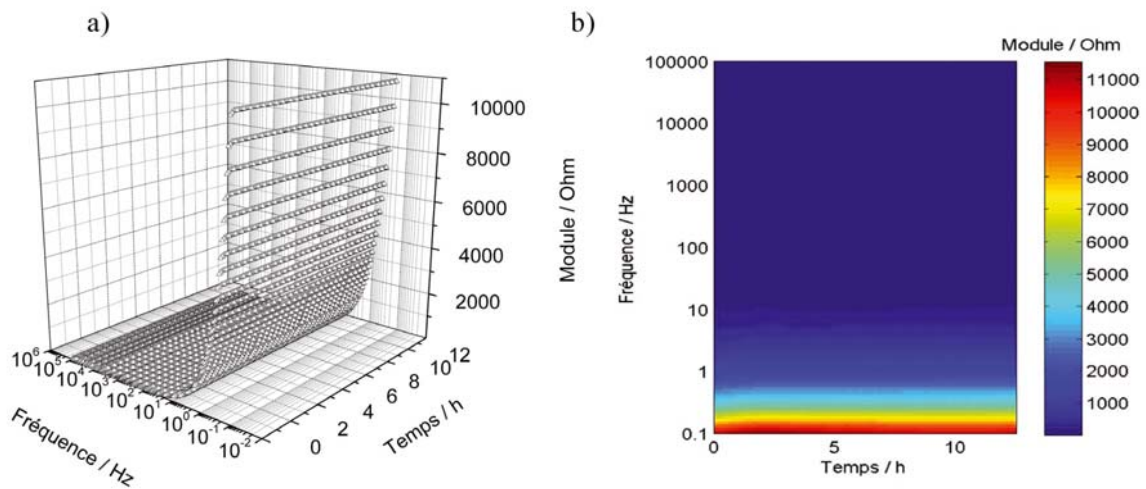

*Figure 2 : a) Evolution de la moyenne du module de 3 essais de SIE dans l'eau désionisée en fonction de la fréquence et du temps ; b) Evolution de la moyenne du module de 3 essais de SIE dans l'eau désionisée, projection dans le plan temps-fréquence.*

*Figure 2: a) Evolution of the mean module for 3 SIE tests in desionised water according to the frequency and the time; b) Evolution of the mean module for 3 SIE tests in desionised water, projection in the time-frequency plane.*

Il est ainsi proposé de représenter l'évolution de la moyenne des modules et des phases, ainsi que le coefficient de variation (écart type divisé par la moyenne) de trois expériences effectuées dans les mêmes conditions. La comparaison sera effectuée entre les 6 solutions évoquées dans le Tableau 2. Ensuite, les expériences effectuées dans l'eau désionisée seront considérées comme la référence ; les différences entre les moyennes des modules des solutions 2 à 6 et la moyenne des modules de la solution 1, la référence, seront représentées pour comparer les évolutions au cours du temps.

*3.2 Influence de la solution d'étude sur l'évolution du module et de la phase*

La Figure 3 présente les variations des moyennes des modules pour chaque solution étudiée ainsi que les coefficients de variation.

On peut constater que les valeurs du module pour l'eau désionisée, la solution de Ringer et la solution de PBS sont quasiment égales. Les expériences, menées dans la solution de PBS + Alb, mettent en évidence les valeurs maximales de module les plus faibles, 8000 Ω pour cette solution, 9000 Ω pour la solution PBS + Sérum + $NaN_3$, et supérieure ou égale à 10000 Ω pour les autres solutions. On peut supposer que l'alliage de titane est le moins protégé dans la solution PBS + Alb. L'albumine favoriserait ainsi plus la corrosion de l'alliage de titane. Il est intéressant de constater que le comportement de l'alliage dans la solution contenant du PBS + Alb, solution 4, et celle contenant du sérum bovin, solution 5, n'est pas le même.

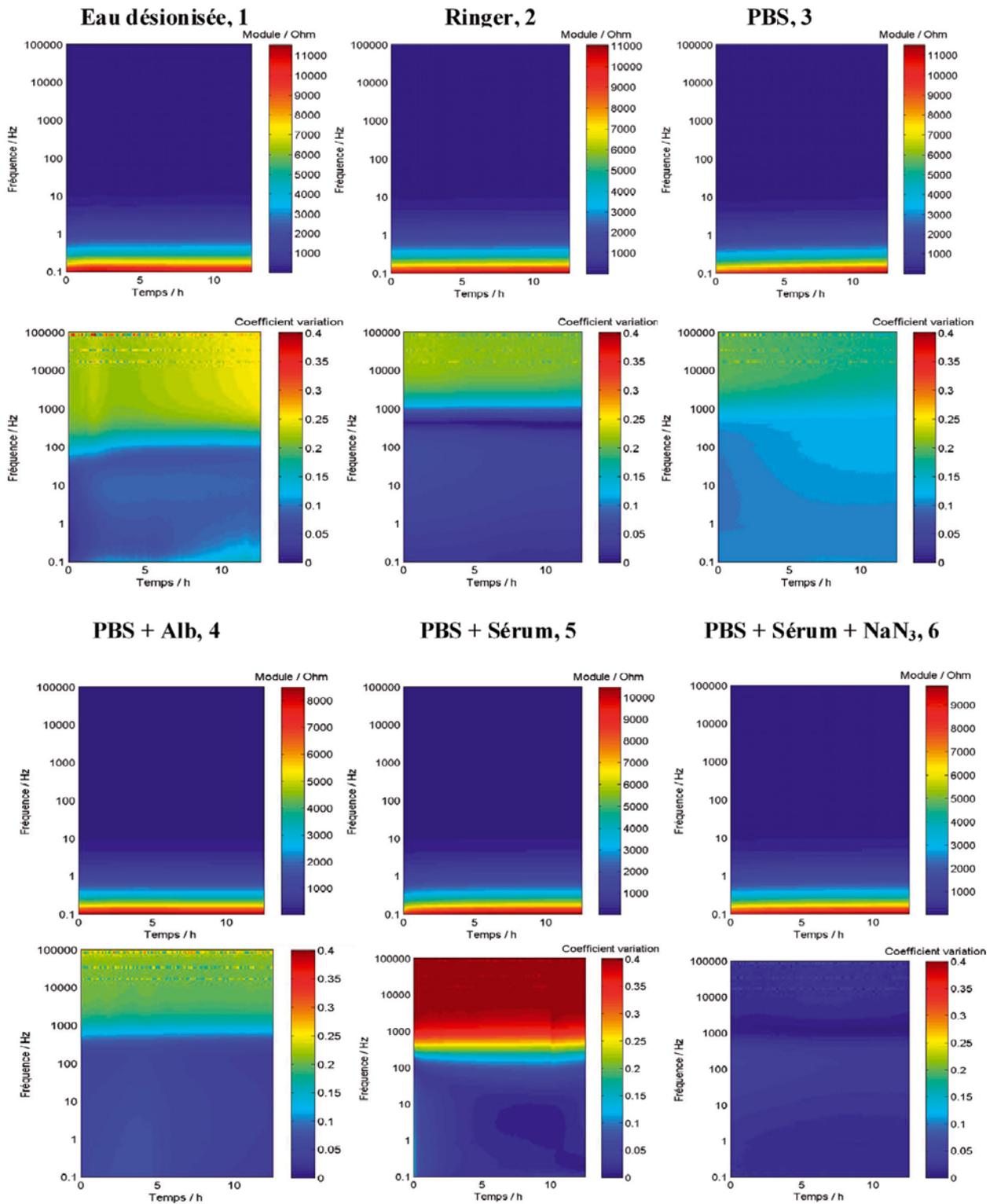

*Figure 3 : Représentation dans le plan temps-fréquence de la moyenne des modules et du coefficient de variation dans les 6 solutions d'étude*

*Figure 3: Time-frequency graphs of the mean module and the coefficient of variation in the 6 investigated solutions.*

Le coefficient de variation obtenu pour la solution 5, jusqu'à 40 %, montre une très mauvaise reproductibilité. Ce résultat peut être interprété comme l'influence de la compétition entre l'adsorption des différents sels contenus dans le sérum bovin et des différentes protéines.

Le coefficient de variation obtenu pour la solution 6, contenant de l'azoture de sodium, est le plus faible, i.e. inférieur à 4 %. Cet antioxydant est ajouté dans les solutions de protéines pour éviter leur dégradation à l'air.

L'analyse des coefficients de variation des expériences issues de l'eau désionisée et de la solution PBS + Alb montre une irrégularité des résultats entre 100000 et 10000 Hz. Ce mode de représentation met en évidence les perturbations du signal à hautes fréquences.

La Figure 4 présente les différences entre les moyennes des modules des solutions de Ringer, PBS, PBS + Alb, PBS + Sérum et PBS + Sérum + $NaN_3$ et la moyenne des modules de l'eau distillée. Ces représentations temps-fréquence peuvent être interprétées en termes de signatures graphiques. Tout d'abord, les valeurs des modules sont toutes inférieures à celle de l'eau désionisée. En conséquence, cette solution est la plus protectrice pour le Ti-6Al-4V.

On peut remarquer que le graphe Ringer-eau est proche de PBS-eau, PBS + Sérum-eau et PBS + Sérum + $NaN_3$-eau. Une différence notable des modules est mise en évidence pour la combinaison PBS + Alb-eau, dans le domaine 1 et 0,1 Hz. On peut conclure que la réponse électrochimique de la surface de Ti-6Al-4V, en présence de PBS + Albumine, est bien différente. Il nous est impossible, à ce stade des investigations, de connaître la composition des couches adsorbées. La représentation des phases permettrait de discuter de l'apparition d'une deuxième boucle capacitive, caractéristique d'une deuxième couche distincte à la surface de l'alliage. Cette discussion sera proposée ultérieurement.

En fonction des solutions utilisées, nous proposons d'émettre des hypothèses sur la constitution des couches jouant un rôle sur l'impédance électrochimique. Pour toutes les solutions, sauf PBS + Albumine et PBS + Sérum + $NaN_3$, les représentations Temps-Fréquence de l'évolution des différences des modules entre les solutions d'étude et l'eau désionisée montrent que les mesures sont stables, écart inférieur à 500 Ω, après 14 heures d'immersion (2 heures à potentiel libre et 12 heures de SIE).

Les résultats des expériences PBS + Albumine et PBS + Sérum + $NaN_3$ montrent que les différences de module entre ces solutions et l'eau désionisée sont supérieures à 500 Ω, entre 1 et 0,1 Hz. Les valeurs de module sont ainsi nettement inférieures à celles de l'eau. En conséquence, les valeurs de résistance, $R_{hf}$, de polarisation sont inférieures. Finalement, l'alliage de Ti-6Al-4V serait le moins protégé dans ces deux dernières solutions, comparativement à l'eau distillée. Pour vérifier l'adsorption de l'albumine, il conviendrait de mener des investigations grâce à la spectrométrie de photoélectrons, XPS, pour avoir accès à la composition de la surface.

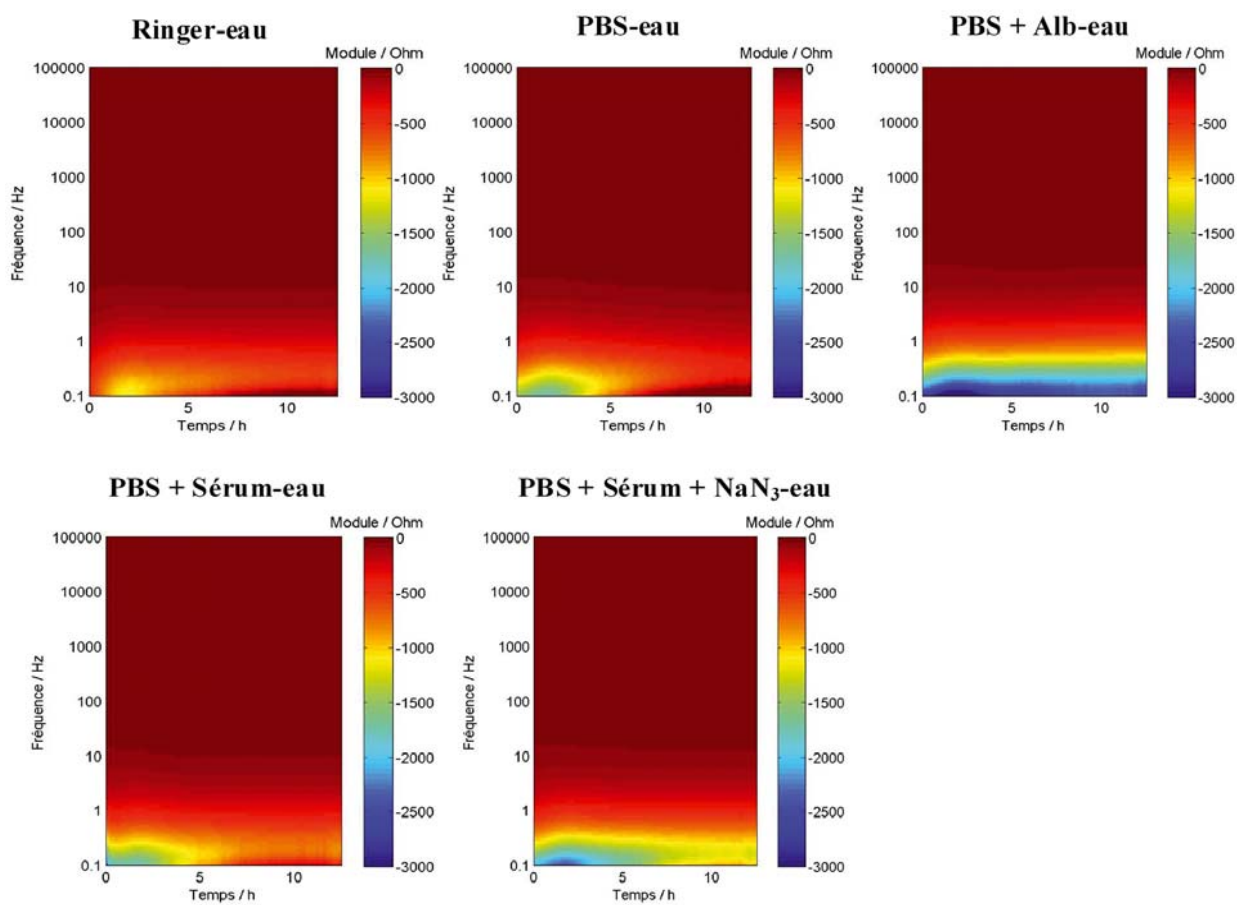

*Figure 4 : Représentation dans le plan temps-fréquence de la différence des moyennes des modules entre cinq solutions et l'eau désionisée servant de référence.*

*Figure 4: Time-frequency graphs of the difference of mean modules between 5 solutions and the reference, desionised water.*

Il est intéressant de noter que les phénomènes de corrosion avec une solution PBS contenant de l'albumine seront différents de ceux avec la même solution tampon contenant du sérum bovin. Enfin, la présence de l'antioxydant, $NaN_3$, dans le sérum bovin modifie notablement le comportement de la surface de l'alliage de titane comparativement au sérum bovin seul.

A l'issue de ces résultats dans le plan temps-fréquence, il apparaît que dans la solution de PBS + Sérum, solution 5, le coefficient de variation est voisin de 40 % (Figure 3), prouvant ainsi que des mécanismes non reproductibles se produisent. Ensuite, la différence des moyennes des modules entre la solution PBS + Alb et l'eau désionisée montre que la signature graphique dans le plan temps-fréquence est particulièrement significative.

La Figure 5 présente les images MEB (microscope JSM 840) caractéristiques des surfaces de Ti-6Al-4V ayant été immergées dans les solutions considérées dans cette étude. Deux images se distinguent i.e. PBS + Alb (image d) et PBS + Sérum (image e). Le pic de carbone élevé relevé dans un spectre EDS montre que des protéines sont adsorbées en plaques à la surface du titane (cette analyse pourra être complétée par une analyse XPS). Ces dépôts ne recouvrent pas toute la surface, expliquant le manque de reproductibilité. Il est aussi très intéressant de noter que la surface de l'échantillon ayant été en contact avec la solution PBS + Sérum + $NaN_3$ ne présente pas de plaques d'adsorption de protéines. Cette observation corrobore les résultats obtenus Figure 3, montrant que l'évolution du module dans cette solution, est très proche de celle obtenue dans les solutions 1, 2 et 3, i.e. sans protéine.

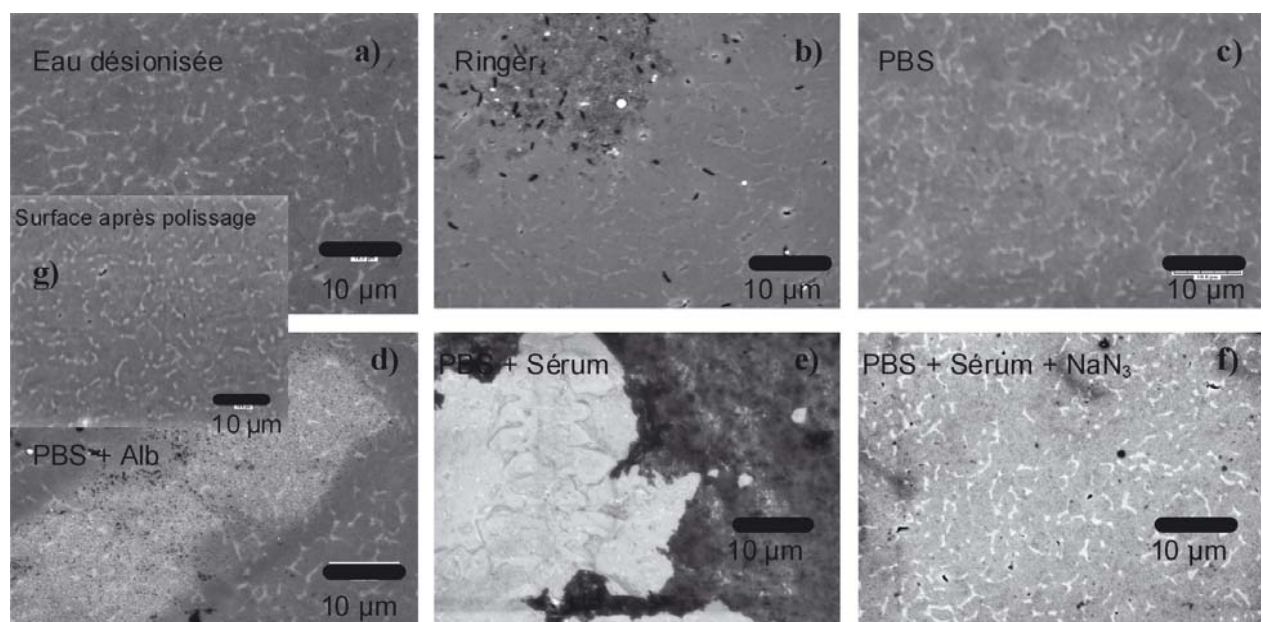

*Figure 5 : Images par Microscopie Electronique à Balayage des surfaces de Ti-6Al-4V après immersion a) dans l'eau désionisée, b) dans la solution de Ringer, c) dans la solution de PBS, d) dans la solution de PBS + Alb, e) dans la solution de PBS + Sérum et f) dans la solution PBS + Sérum + $NaN_3$ ; la durée d'immersion des échantillons est égale à 40 heures. g) Surface de l'alliage de titane après l'étape de polissage.*

*Figure5: Scanning Electron Microscopy images of Ti-6Al-4V after experiments in a) desionised water, b) Ringer solution, c) PBS solution, d) PBS + Alb solution, e) PBS + serum solution, f) PBS + serum + $NaN_3$; the immersion time is of 40 hours. g) Titanium alloy surface after the polishing step.*

### 4. Conclusions

La spectroscopie d'impédance électrochimique et les représentations temps-fréquence proposées en découlant, permettent d'obtenir des informations sur le comportement passif d'un matériau. Le comportement de l'alliage Ti-6Al-4V, utilisé dans le domaine des implants orthopédiques et dentaires, a été étudié dans différentes solutions simulant le liquide physiologique. Une représentation temps-fréquence a été proposée pour suivre l'évolution cinétique du module, du coefficient de variation, 3 essais par solution, et de la différence entre les moyennes des modules et la moyenne des modules de l'eau désionisée.

Tout d'abord, il ressort que le module le plus faible a été mesuré dans la solution PBS + Alb, i.e. la moins protectrice. Le coefficient de variation du module, dans la solution de PBS + Sérum, est le plus élevé, environ 40 % ; la reproductibilité des mesures est mise en défaut. D'après les images MEB, l'adsorption des protéines n'est pas homogène, ce qui confirme les problèmes de reproductibilité. Le coefficient de variation du module, dans la solution de PBS + Sérum + $NaN_3$, est le plus faible, environ 4 %. D'après les images MEB, aucune plaque de protéines d'adsorption n'a été visible. L'antioxydant empêcherait l'adsorption de plaques de protéines à la surface de cet alliage.

L'analyse des différences entre les moyennes des modules des solutions 2, 3, 4, 5, 6 et la moyenne des modules de la solution d'eau désionisée met en évidence deux types de comportement. Le premier est relatif aux solutions Ringer, PBS, PBS + Sérum et le deuxième correspond aux solutions PBS + Alb et PBS + Sérum + $NaN_3$. Le module, dans ces deux dernières solutions est plus faible que celui mesuré dans l'eau désionisée. De plus, la représentation temps-fréquence est vraiment différente du premier comportement. L'influence des espèces adsorbées sur le comportement électrochimique de l'alliage de Ti-6Al-4V est ainsi nettement différente entre les deux groupes de solution. Les investigations comprenant des solutions contenant de l'albumine et du sérum bovin ne sont pas comparables car la réponse électrochimique est différente après 14 heures d'immersion. La solution PBS + albumine engendre une résistance à la corrosion moindre que la solution PBS + sérum bovin. De plus l'antioxydant, $NaN_3$, joue un rôle important et tend à diminuer la passivité de l'alliage. Enfin, le comportement électrochimique de l'alliage Ti-6Al-4V, en solution de Ringer, est proche de celui en solutions PBS et PBS + sérum bovin ; les protéines du sérum ne semblent pas avoir d'influence. Les protéines autres que l'albumine pourraient expliquer la différence de comportement.

Des investigations sont en cours pour étudier le comportement de deux autres biomatériaux métalliques, l'acier inoxydable 316L et le Co-Cr-Mo. De plus, des images AFM et des analyses XPS permettront de caractériser plus finement l'état de surface et la couche de protéines adsorbées, le cas échéant.

**Références :**

[1] J.B. Park, R.S. Lakes, Biomaterials an Introduction, 2$^{nd}$ ed., Plenum Press, New York, 1992, 83-88.
[2] A. WE. Hodgson, S. Kurz, S. Virtanen, V. Fervel, C.-O.A. Olsson, S. Mischler, Electrochimica Acta 49 (2004) 2167
[3] A. Ouerd, C. Alemany-Dumont, B. Normand, S. Szunerits, Electrochimica Acta 53 (2008) 4461
[4] A.I. Munoz, S. Mischler, Journal of The Electrochemical Society 154 (2007) C562
[5] C. Valero, A.I. Munoz, Corrosion Science 50 (2008) 1954
[6] D. Dowson, V. Wright, An introduction to the biomechanics of joints and joint replacement, Mechanical Engineering Publications, London, 1981, 174
[7] J. D. Bronzino, The Biomedical Engineering Handbook, CRC Press, Boca Raton Florida, 1995, 540-542
[8] C. Liu, Q. Bi, A. Matthews, Surface and Coatings Technology 163-164 (2003) 597
[9] A. Kirbs, R. Lange, B. Nebe, R. Rychly, A. Baumann, H.-G. Neumann, U. Beck, Materials Science and Engineering C23 (2003) 425
[10] R.W-W. Hsu, C-C. Yang, C-A. Huang, Y-S. Chen, Materials Science and Engineering A380 (2004) 100
[11] M.A. Khan, R.L. Williams, D.F. Williams, Biomaterials 20 (1999) 631
[12] J. Pan, D. Thierry, C. Leygraf, Electrochimica Acta 41 (1996) 1143
[13] S. Tamilselvi, V. Raman, N. Rajendran, Electrochimica Acta 52 (2006) 839
[14] N. Zaveri, M. Mahapatra, A. Deceuster, Y. Peng, L. Li, A. Zhou, electrochimica Acta 53 (2008) 5022
[15] A. Ouerd, C. Alemany-Dumont, G. Berthomé, B. Normand, S. Szunerits, Journal of The Electrochemical Society 154 (2007) C593
[16] H. Yan, L. Xiaoying, M. Jingwu, H. Nan, Applied surface Science 255 (2008) 257
[17] K. Cai, M. Frant, J. Bossert, G. Hildebrand, K. Liefeith, K. D. Jandt, Colloids and Surfaces B: Biointerfaces 50 (2006) 1
[18] V. Payet, S. Brunner, A. Galtayries, I. Frateur, P. Marcus, Surf. Interface Anal. 40 (2008) 215
[19] J. Geringer, B. Normand, R. Diemiaszonek, C. Alemany-Dumont, N. Mary, Matériaux et Techniques 95 (2007) 417